\documentclass[11pt]{article}
\usepackage{lineno,hyperref}
\modulolinenumbers[5]
\usepackage{stmaryrd}
\usepackage{subfigure}
\usepackage{color} 
\usepackage{amsmath}
\usepackage{bm}
\usepackage{bbm}
\usepackage{mathrsfs}
\usepackage{graphicx}
\usepackage{caption}
\usepackage{epsfig}
\usepackage{amsfonts}
\usepackage{amsmath}
\usepackage{graphicx}
\usepackage{psfrag}
\usepackage{placeins}
\graphicspath{{./figures/}}
\def\Bvarphi{\mbox{\boldmath$\varphi$}}

\def\bF{\mbox{\boldmath$ F$}}

\def\bN{\mbox{\boldmath$ N$}}

\def\bP{\mbox{\boldmath$ P$}}

\def\bX{\mbox{\boldmath$ X$}}

\def\be{\mbox{\boldmath$ e$}}

\def\bn{\mbox{\boldmath$ n$}}

\def\bu{\mbox{\boldmath$ u$}}
\def\bv{\mbox{\boldmath$ v$}}

\def\bx{\mbox{\boldmath$ x$}}

\begin{document}
\title{A computational study of the mechanisms of growth-driven folding patterns on shells, with application to the developing brain}
\author{S.N.~Verner\thanks{Department of Mechanical Engineering, University of Michigan}, K.~Garikipati\thanks{Departments of Mechanical Engineering and Mathematics, University of Michigan}\thanks{corresponding author: {\tt krishna@umich.edu}}}
\maketitle
\begin{abstract} We consider the mechanisms by which folds, or sulci (troughs) and gyri (crests), develop in the brain. This feature, common to many \emph{gyrencephalic} species including humans, has attracted recent attention from soft matter physicists. It occurs due to inhomogeneous, and predominantly tangential, growth of the cortex, which causes circumferential compression, leading to a bifurcation of the solution path to a folded configuration. The problem can be framed as one of buckling in the regime of linearized elasticity. However, the brain is a very soft solid, which is subject to large strains due to inhomogeneous growth. As a consequence, the morphomechanics of the developing brain demonstrates an extensive post-bifurcation regime. Nonlinear elasticity studies of growth-driven brain folding have established the conditions necessary for the onset of folding, and for its progression to configurations broadly resembling gyrencephalic brains. The reference, unfolded, configurations in these treatments have a high degree of symmetry--typically, ellipsoidal. Depending on the boundary conditions, the folded configurations have symmetric or anti-symmetric patterns. However, these configurations do not approximate the actual morphology of, e.g., human brains, which display unsymmetric folding. More importantly, from a neurodevelopmental standpoint, many of the unsymmetric sulci and gyri are notably robust in their locations. Here, we initiate studies on the physical mechanisms and geometry that control the development of primary sulci and gyri. In this preliminary communication we carry out computations with idealized geometries, boundary conditions and parameters, seeking a pattern resembling one of the first folds to form: the Central Sulcus.
\end{abstract}
\section*{Keywords} morphology; patterning; cortical folding; elasticity; bifurcation

\section{Introduction}

Folding, or sulcification and gyrification, of the brain is common in mammals including primates, cetaceans, pachyderms and ungulates. Folds form in the  cortical layer of grey matter, and in species such as humans that demonstrate pronounced gyrencephaly, the sulci can be significantly deeper than the cortical thickness. From a neurophysiological point of view, a folded cortex confers a cognitive advantage by increasing the surface area enclosed within the skull, translating to greater capacity for intelligence. Human brains in a nonpathological state have a gyrification index (ratio of actual surface area to the surface area of an enveloping surface) approaching 2.55 \cite{Zilles1988}. Neurodevelopmental pathologies are associated with significant departures from this value. In humans, polymicrogyria (shallow, more frequent folding) is associated with  developmental delays and epilepsy \cite{Barkovich2010}. Pachygyria (shallow, less frequent and flatter folds) can cause seizures, mental retardation and in rare cases, mania \cite{Seshadri2016}. Lissencephaly (abscence of folds) is linked to abnormal EEG patterns, mental retardation and agitation, and manifests in under-developed social skills \cite{Landrieu1998}. 

 \indent Fetal MRI data indicate that the human brain is almost perfectly smooth until 24 weeks of gestation \cite{Habas2012, Gholipour2014, Gholipour2017}, from which stage gyrification proceeds until well after birth. Therefore, there is a clear neurophysiological motivation to understand the physics governing cortical folding and the conditions for normal or pathological cortical folding. 
 
\indent There have been competing hypotheses for this phenomenon. Most prominent have been (a) the axonal tension model of cortical folding under forces imposed by interconnected neurons \cite{vanEssen1997}---a theory in turn challenged by (b) the principle of inhomogeneous growth of the cortical layer in which circumferential compression due to growth causes an elastic buckling bifurcation, and extreme strains lead to highly folded structures in the post-bifurcation regime. Studies of cutting followed by elastic relaxation on ferret brains established that axonal tension does not cause folding, while computational studies strongly suggested that inhomogeneous growth does \cite{Xu2010}. Bayly et al \cite{Bayly2013} explained gyrification patterns by analytic and computational studies based on inhomogeneous growth, and Tallinen et al \cite{Tallinen2016} used experiments in a surrogate, polymeric gel model combined with nonlinear finite element computations to further support the inhomogenous growth theory.\footnote{Albeit, solved as elastic unloading from the folded configuration with first-order dynamics added to numerically stabilize the system against bifurcations.} 

\indent Mismatched elastic moduli between a thin elastic layer and an underlying substrate are common in many non-biological thin film applications \cite{Huang2004}. Such stiffness contrast also is a feature that may control the patterns of wrinkling of fruit and vegetable skins \cite{Yin2008}. However, it is not essential to brain folding  \cite{Prange2002, Chatelin2010, vanDommelen2010}; the Young's Modulus of cortical grey matter and of the white matter underlying it are of the same order of magnitude \cite{Budday2017_2}.

\indent There is now a sizeable literature \cite{Bayly2013,Tallinen2016, Budday2017,Budday2014, Budday2017_2,Budday2015, Goriely2015, Hutchinson1975,Tallinen2013} seeking to explain aspects of brain folding by inhomogeneous growth in linearized and, more appropriately, nonlinear elasticity. Some of this literature draws from linearized buckling of beams and plates \cite{Goriely2015, Hutchinson1975, Budday2015_2}, but much of the computational work is based on finite strains, and operates in the post-bifurcation regime. This work has shed light on the mechanical conditions governing the development of the organ-wide pathologies of polymicrogyria, pachygyria and lissencephaly \cite{Budday2014,Goriely2015,Budday2015}. However, the precise form of the folded cortex is important beyond its implications for these pathologies. In humans and other gyrencephalic species, the normally developed brain does not fold into perfectly symmetric or antisymmetric mode shapes that may be expected from elastic buckling and post-bifurcation straining on reference configurations of high symmetry. Primary sulci and gyri--the early forming, prominent folds--are not localized into either symmetric or anti-symmetric modes of folding \cite{Habas2012, Paus1996}. Studies of the sequence of normal formation of primary sulci and gyri, however, are currently lacking.

Here, we initiate studies on the geometry and physical mechanisms that, governed by the phenomenology of inhomogeneous growth, lead to primary sulci and gyri in the normally developed human brain. In this first communication, we vary (a) geometries guided by quantitative data from anatomical measurements, and (b) mechanisms of cell accumulation by local proliferation, and by migration. Our goal is to reproduce a pattern that suggests the incipient Central Sulcus (Figure \ref{fig:Potato}a). Apart from its location, which is roughly in the coronal plane, and its orientation, which is close to vertical, this target is recognized qualitatively rather than quantitatively in this preliminary computational study. We exploit the smoothness of the 24 week-old fetal brain \cite{Habas2012, Gholipour2014, Gholipour2017}, a convenient reference configuration, relative to which we consider growth.

\emph{Most previous studies have reduced the problem to one of local, inhomogeneous growth controlled by a time- or load-dependent scalar parameter \cite{Bayly2013,Tallinen2016, Budday2017,Budday2014, Budday2017_2,Budday2015,Tallinen2013}. Effectively, this addresses only the mechanism of local cell proliferation. In contrast, we also pay attention to the developmental processes by which neurons arise near the ventricles and migrate outward to the cortex \cite{Jin2003, Sun2014}. There, they intercalate circumferentially, causing tangential growth \cite{Ronan2014} in the two-dimensional surface manifold that is the cortical layer. We use the advection-diffusion-reaction equation to model cell migration and proliferation, and couple it to a local model of tangential growth.}

\indent Our treatment begins with the governing and constitutive equations in Section \ref{sec:modelgoveqns}. The computational framework is briefly presented in Section \ref{sec:compframework}, followed by studies of the effects of: geometry (Section \ref{sec:geom}), mechanisms of cell migration (Section \ref{sec:cellmotion}) and cortical thickness  (Section \ref{sec:corticalthick}). The role that energy variations play in the development of bifurcations is studied in Section \ref{sec:energy}. Closing remarks appear in Section \ref{sec:concl}.

\section{Model and Governing Equations}
\label{sec:modelgoveqns}

\begin{figure}[h]
\includegraphics[scale = 0.15]{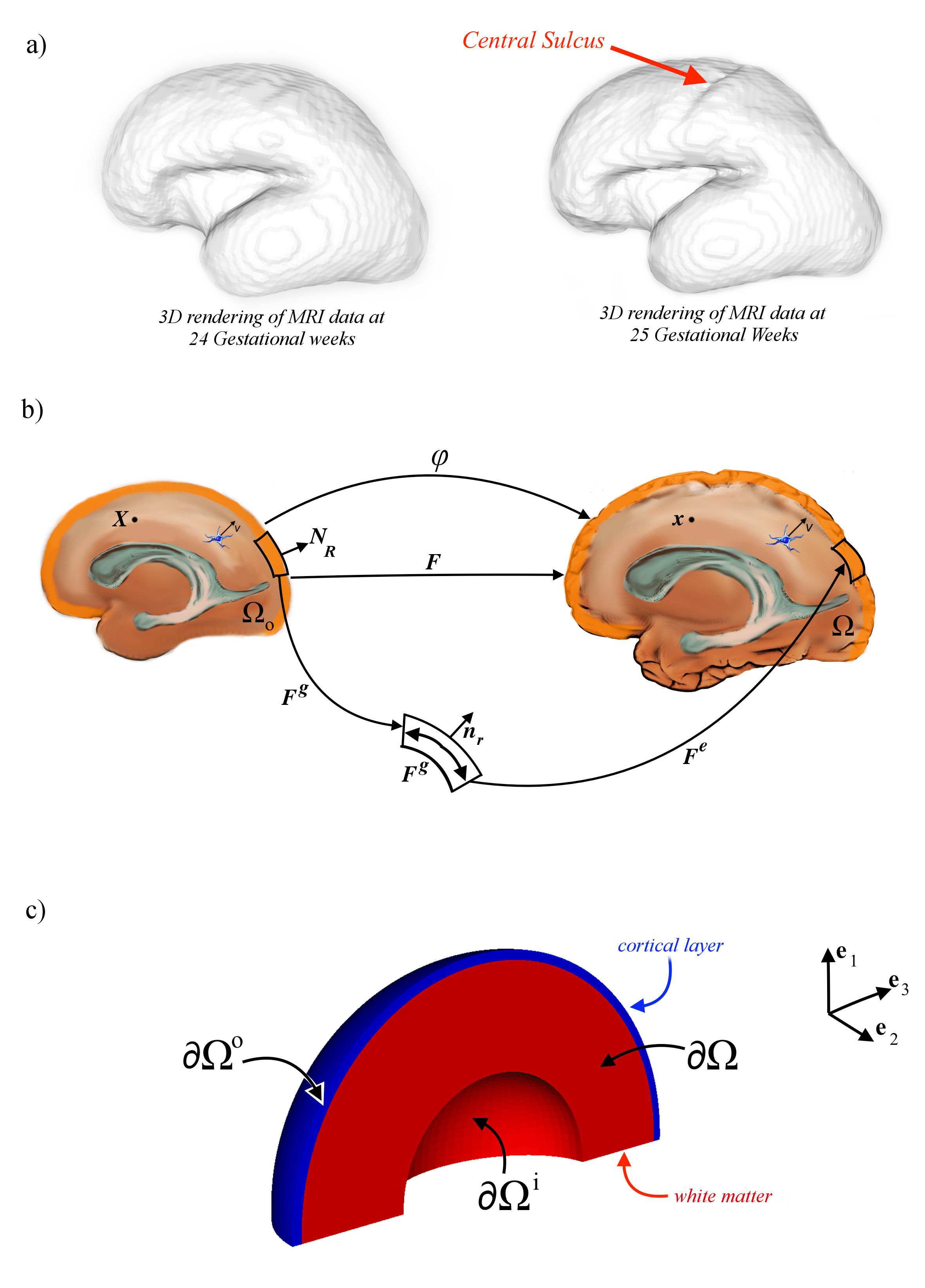}
\centering
\caption{(a) 3D rendering of fetal MRI data at 24 and 25 Gestational Weeks, b) Illustration of the brain as a deforming (growing) continuum body in reference and deformed configurations and (c) illustration of domain boundaries on a mathematically idealized deformed configuration $\Omega$.}
\label{fig:Potato}
\end{figure}

We adopt the classical formulation of continuum mechanics. The reference configuration representing the smooth, fetal brain is denoted by $\Omega_0$. Reference positions of material points are vectors $\bX \in \Omega_0 \subset\mathbb{R}^3$, and the displacement field vector is $\bu \in \mathbb{R}^3$. Points in the deformed (and grown) configuration, $\Omega$, are labelled $\bx = {\Bvarphi}(\bX) = \bX + \bu$. The deformation gradient tensor is $\bF = \bold{1} + \partial\bu/\partial\bX$, where $\boldmath{1}$ is the second-order isotropic tensor. Figure \ref{fig:Potato}a illustrates these kinematics and a few other key aspects of the treatment. Inhomogeneous growth is modelled by the multiplicative, \emph{elasto-growth} decomposition $\bF = \bF^\text{e}\bF^\text{g}$. Denoting the cell concentration in $\Omega$ by $c$, tangential growth in the cortical layer is written as 

\begin{subequations}
\begin{align}
    \bF^\text{g}(c(\bX)) &=
    \left\{\begin{array}{cl}
    \frac{1}{2-f(c)}\left(\boldmath{1} - (f(c)-1)\bN \otimes \bN\right),& \bX\in \text{cortical layer}\\
    \boldmath{1},&\bX \notin\text{cortical layer}
    \end{array}
    \right.\label{Eq:Fg}\\
    f(c) &=
    \left\{\begin{array}{cc}
    1, & c \leq c_\text{cr}\\
    \frac{c}{c_\text{cr}}, & c > c_\text{cr}
    \end{array}
    \right.
    \label{Eq:fc}
\end{align}
\end{subequations}
with $\bN$ representing the surface normals on $\partial\Omega_0$, respectively. The form of $\bF^\text{g}$ in Equation (\ref{Eq:Fg}) ensures that cell intercalation-driven tangential growth occurs only in the cortex, and within the cortical tangent plane. The form of $f(c)$ in Equation (\ref{Eq:fc}) ensures that tangential expansion occurs only after the cell concentration in the cortex has exceeded the threshold of $c_\text{cr}$, thus modelling the effect of free volume. We use $c_\text{cr}(\bx) = c(\Bvarphi(\bX),0)$, the initial concentration.

We consider hemispherical and hemi-ellipsoidal reference configurations, $\Omega_0$, with cortical layers of varying thicknesses, forming thin shells of grey matter resting on elastic foundations of white matter in each case. The white matter is itself a thick shell with the inner surface, $\partial\Omega^\text{i}$ representing the ventricles (Figure \ref{fig:Potato}c). Since the time scales of growth are much greater than the intrinsic viscoelastic relaxation times of the soft, jelly-like brain, its constitutive response is modelled by an elastically compressible, neo-Hookean strain energy density expressed as a function of $\bF^\text{e}$,

\begin{subequations}
\begin{align}
\bF^\text{e}(c) &= \bF\left(\bF^\text{g}(c)\right)^{-1},\label{Eq:growthFeFg}\\
W(\bF^\text{e}) &=  \frac{1}{4}\lambda(\text{det}\bF^{\text{e}^\text{T}}\bF^\text{e} -1) - \frac{1}{2}(\frac{1}{2}\lambda +
  \mu)(\log\text{det}\bF^{\text{e}^\text{T}}\bF^\text{e}) + \frac{1}{2}\mu(\bF^\text{e}\colon\bF^\text{e} - 3),\label{Eq:neohookean}
\end{align}
\end{subequations}
where $\lambda$ and $\mu$ are Lam\'{e} parameters. The first Piola-Kirchhoff stress $\bP$, and its governing quasistatic equilibrium equation are,

\begin{subequations}
\begin{align}
\bP &= \frac{\partial W}{\partial\bF^\text{e}}\label{Eq:P}\\
    \text{Div}\bP &= \boldmath{0},\quad \text{in}\; \Omega_0.
    \label{Eq:equil}
\end{align}
\end{subequations}

Neuronal migration and production are modelled by an advection-diffusion-reaction equation (\ref{Eq:cell_motion}) written on the deformed and grown configuration, $\Omega$:

\begin{equation}
    \frac{\partial c}{\partial t} = D \nabla^2 c - \bv\cdot\nabla c + R,\quad \text{in}\;\Omega,
    \label{Eq:cell_motion}
\end{equation}
where $D$ is an effective diffusivity modelling random cell migration, $\bv$ is a directed migration velocity, the reaction term, $R$, models cell proliferation, and $\nabla$ is the spatial gradient with respect to $\bx$. The parameters used in our computations are summarized in Table \ref{Tbl:table1}. Near elastic incompressibility is modelled by the ratio of Lam\'{e} parameters, which corresponds to a Poisson ratio $\nu = 0.49$ in the regime of linearized elasticity. The scaling constant $v^\text{c}$ gives the migration velocity's magnitude. The dynamic quantities $D$ and $v^\text{c}$ have been scaled up in magnitude relative to physiological values in order to speed up our computations.

\begin{table}[h]
\caption{Model parameters}
\begin{center}
\begin{tabular}{| c| c|}
\hline
\textit{Parameter} & \textit{Value} \\ \hline
 Diffusivity ($D$) & $0.1\;\text{mm}^2\cdot \text{s}^{-1}$\\
 Lam\'{e} parameter $\lambda$ & $8.2\times10^4\; \text{Pa}$ \\
 Lam\'{e} parameter $\mu$ & $1.67\times 10^3\; \text{Pa}$ \\
 Velocity constant ($v^\text{c}$) & 0.1 \; $\text{mm}\cdot \text{s}^{-1}$\\
 Cellular proliferation constant ($R$) & 0.1 $\text{s}^{-1}$ \\
 Outer Radius of hemispherical brain ($R_\text{o}$) &  20 mm\\
 Inner Radius of hemispherical brain ($r_\text{i})$ & 10 mm\\ \hline
\end{tabular}
\end{center}
\label{Tbl:table1}
\end{table}

\subsection{Initial and Boundary Conditions} 
Working with a non-dimensional cell concentration, we impose initial conditions
\begin{equation}
    c(\bx,0) =
    \left\{\begin{array}{cc}
         1.0& \bx \in\partial\Omega^\text{i} \\
         0.5& \bx \notin\partial\Omega^\text{i}
    \end{array}\right.
    \label{Eq:IC}
\end{equation}
The boundary conditions on growth-driven mechanics are,

\begin{subequations}
\begin{align}
\bu(\bX) &= \boldmath{0},\;\text{for}\; \bX\in \partial\Omega_{0}^\text{i}\label{Eq:MechBCa}\\
u_1(\bX) &= 0,\;\text{for}\; X_1 = 0\label{Eq:MechBCb}\\
\bP\bN &= \boldmath{0},\;\text{for}\; \bX\in \partial\Omega\backslash\partial\Omega_{0}^\text{i}\label{Eq:MechBCc}
\end{align}
\end{subequations}
and on the advection-diffusion-reaction of cells:
\begin{subequations}
\begin{align}
c(\bx,t) &= 1.0,\;\text{for}\; \bx\in \partial\Omega^\text{i}\label{Eq:TranspBCa}\\
(-D\nabla c + c\bv)\cdot\bn &= 0,\;\text{for}\; \bx\in \partial\Omega\backslash\partial\Omega^\text{i}\label{Eq:TranspBCb}
\end{align}
\end{subequations}
representing cell birth on $\partial\Omega^\text{i}$. The distinct boundaries have been delineated in Figure \ref{fig:Potato}c.

The initial conditions (\ref{Eq:IC}) and boundary conditions (\ref{Eq:TranspBCa}-\ref{Eq:TranspBCb}) applied to Equation (\ref{Eq:cell_motion}) drive the neuronal population from the ventricles bounded by $\partial\Omega^\text{i}$ toward the cortical surface bounded by $\partial\Omega^\text{o}$. In the cortex, cell intercalation drives tangential growth, creating a compressive circumferential stress that induces a buckling bifurcation, and post-bifurcation straining into folded structures of sulci and gyri. In Sections \ref{sec:geom}-\ref{sec:corticalthick}, we have examined the influences of geometry, mechanism of cell migration, and cortical thickness on patterns of folding, with the goal of identifying a structure that approximates an incipient Central Sulcus.

\section{Computational framework}
\label{sec:compframework}
\subsection{Numerical methods}
The partial differential equations described in Section \ref{sec:modelgoveqns} have been solved by finite element methods, using trilinear hexahedral elements. The backward Euler algorithm has been employed for time integration of the advection-diffusion-reaction equation (\ref{Eq:cell_motion}), and the Streamline Upwind Petrov Galerkin (SUPG) method for its stabilization, which becomes important in the hyperbolic limit. The nonlinear residual equations have been consistently linearized, with Jacobians obtained by automatic differentiation (see Section \ref{sec:software}), and deployed in a direct Newton-Raphson iteration scheme without continuation.
\subsection{Software}
\label{sec:software}
The finite element formulation has been implemented in {\tt C++} by extending a code for general problems of patterning and morphology \cite{Garikipati2017}. This framework uses the {\tt deal.II} open source finite element library \cite{Bangerth2007,Bangerth2016}. Code parallelization is based on {\tt MPI}. Automatic differentiation is implemented using the {\tt Sacado} library from the {\tt Trilinos} software suite. The SuperLU direct solver \cite{Demmel1999} has been used. The code for all numerical examples presented here is available at
{\tt https://github.com/mechanoChem/patternMorph}. Post-processing was carried out in the visualization toolkit {\tt Visit 2.12.0} \cite{HPV:VisIt}. 
{\subsection{Data}
Brain geometries were obtained from the fetal brain atlas developed by Gholipour et al \cite{Gholipour2014, Gholipour2017} and can be downloaded at {\tt http://crl.med.harvard.edu/ \\research/fetal\_brain\_atlas}. The  magnetic resonance imaging (MRI) viewing software {\tt ITK-Snap} \cite{Yushkevich2006} has been used to obtain the geometric parameters, namely aspect ratios and cortical thickness for our finite element models.} 

\section{The influence of geometry: Hemispherical and hemi-ellipsoidal models}
\label{sec:geom}

\begin{figure}[h]
\includegraphics[scale = 0.15]{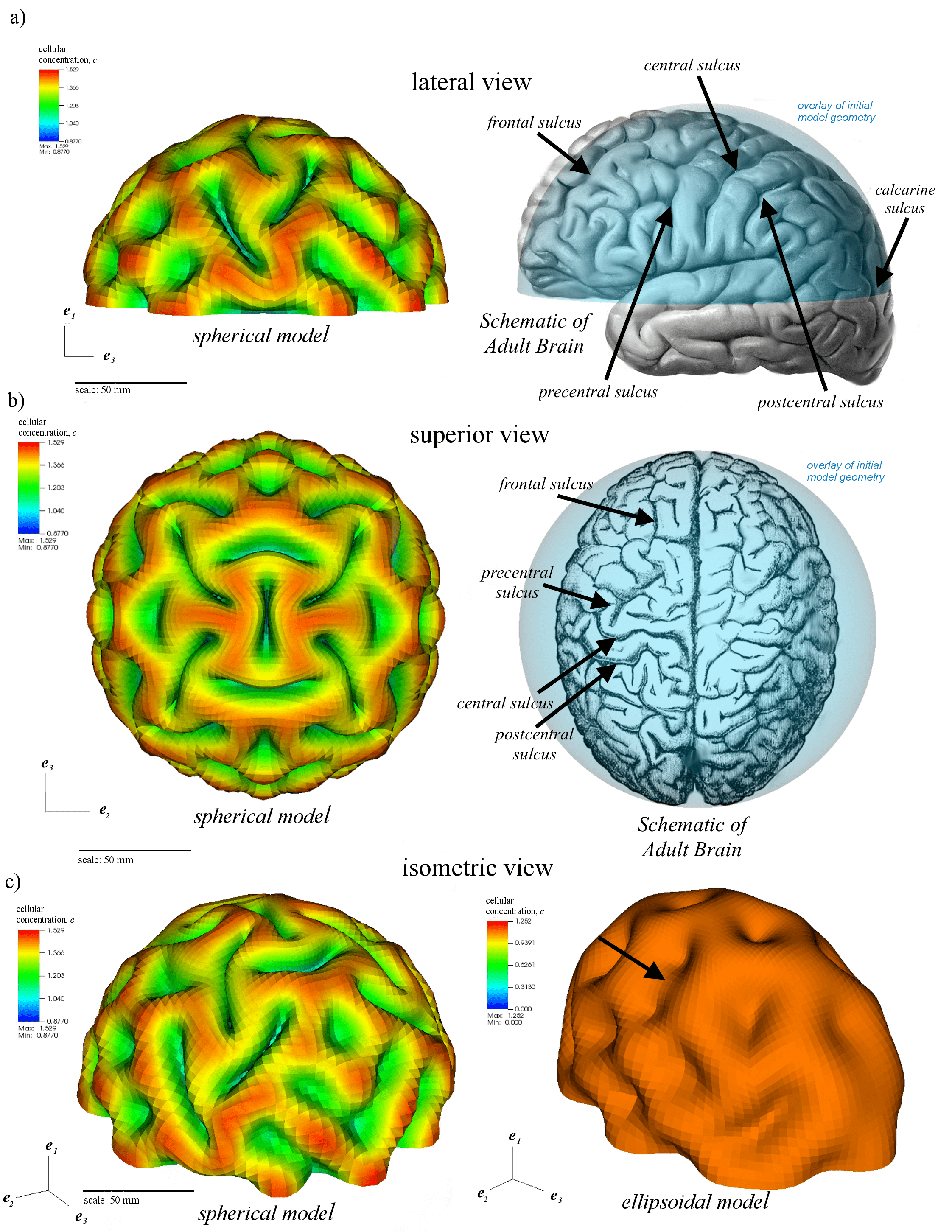}
\centering
\caption{a) Lateral views of hemispherical model of cortical folding, left and schematic of adult human brain (Wikipedia Commons, public domain image) with an overlay of the starting computational geometry in blue, right, b) Superior views of hemispherical model of cortical folding, left and artistic representation of a normal human brain (Wikipedia Commons, public domain image) with an overlay of the starting computational geometry in blue, right, and (c) An isometric view comparison of the hemispherical model, left, and hemi-ellipsoidal model, right.}
\label{fig:geometry}
\end{figure}

We first studied the influence of geometry, starting with a hemispherical approximation of the brain, using an outer radius of 20 mm and an inner radius of 10 mm, shown in Table \ref{Tbl:table1}. With Equations (\ref{Eq:Fg}-\ref{Eq:TranspBCb}), $R = 0$ and other coefficients as in Table \ref{Tbl:table1}, the model drives cells from the ventricular surface $\partial\Omega^\text{i}$ into the cortex, where their accumulation causes buckling, which develops through post-bifurcation deformation into a prominently folded state. On the finite element mesh of the hemispherical geometry, the first elastic bifurcations are located in regions of transitions to  the finest elements. However, the final, post-bifurcated, folded configuration demonstrated insensitivity to these initial perturbations, varying instead with the cortical thickness as reported in the literature \cite{BenAmar2010}. The resulting patterns of sulci and gyri in Figure \ref{fig:geometry}a, b and c (left column) adopt a triangular lattice-like arrangement on the spherical surface, similar to the results of Tallinen et al \cite{Tallinen2013} on flat surfaces. However, this degree of regularity is not seen in the pattern of sulci and gyri in the human brain, schematics of which appear in Figure \ref{fig:geometry}a and b (right column). Specifically, the primary folds, such as the Central Sulcus, Frontal Sulcus, Pre- and Post-central sulcus, the Calcarine Sulcus and others, \cite{Habas2012} are not seen.

In seeking to better model the primary folds, we drew upon the fact that the human brain's shape is better approximated by a hemi-ellipsoid than a hemisphere, as indicated by fetal MRI data \cite{Habas2012,Gholipour2014,Chapman2010,Gholipour2017}. From data in Gholipour et al. at 24 Gestational Weeks \cite{Gholipour2014, Gholipour2017}, the point before sulcification and gyrification, we infer the ratio of the best-approximating ellipsoid's axes to be $1\colon 0.75\colon 1$ along the $\be_1, \be_2, \be_3$ directions. The shortest axis, $\be_2$, is perpendicular to the sagittal plane (as illustrated in Figure \ref{fig:geometry}b, right), and the major axes have lengths 20, 15 and 20 mm for the outer surface and 10, 7.5 and 10 mm for the inner surface.

It is trivial that, in the spherical geometry, a symmetric advective velocity field, $\bv$, is both radial and normal to concentric spheres. In the ellipsoidal geometry, however, a radial field is distinct from one that is normal to concentric surfaces, as is easily checked. Cells migrating by radial or normal advective velocity fields of uniform magnitude collect in non-uniform concentrations on the cortical surface, as is also easily verifiable. These effects of the ellipsoidal geometry are considered in Section \ref{sec:cellmotion}. In order to isolate the effect of spherical/ellipsoidal geometry from the advective velocity, we return to Equation (\ref{Eq:cell_motion}) and set $\bv = \boldmath{0}$ and $D = 0$, with $R$ set to the value in Table \ref{Tbl:table1}. This independence from cell migration in favor of cell proliferation has been adopted by several authors previously \cite{Bayly2013,Tallinen2016,Budday2017,Budday2014,Budday2017_2,Budday2015,Budday2015_2,Budday2015_3,Budday2014_2}. 

The resulting morphology is shown in Figure \ref{fig:geometry}c, right, and compared with the spherical model in Figure \ref{fig:geometry}c, left. Note that both models employ a cortical thickness of 10\%. As demonstrated by the computations, the ellipsoidal geometry generates a structure resembling an incipient Central Sulcus (arrow in Figure \ref{fig:geometry}c, right).

Motivated by this result, the remainder of this communication is focused on the ellipsoidal geometry. On it, we explore the role of mechanisms of cell accumulation by migration or proliferation. 

\section{Neuronal migration mechanisms: The influence of cell velocity distributions}
\label{sec:cellmotion}

As outlined in Section \ref{sec:geom}, if the hemisphere is considered a degenerate hemi-ellispoid, it is clear that there are at least two advective velocity distributions on the ellipsoid that collapse to the radial distribution on the sphere. The influences on cortical folding patterns, due to the corresponding radial and normal velocity distributions for cell advection, are developed in detail in Sections \ref{sec:radialmap} and \ref{sec:normalmap}. In Section \ref{sec:compare}, their influences on formation of the incipient Central Sulcus are compared with the cell proliferation model of Section \ref{sec:geom}. The hemi-ellipsoidal representation of the fetal brain introduced in Section \ref{sec:geom} remains our geometric model for this study of cell velocity fields.
\subsection{Radial advection velocity in the ellipsoidal geometry}
\label{sec:radialmap}

In order to radially orient the advection velocity vector, we first define a unit vector (a direction), $\hat{\bx}^\text{e}$ in Equation (\ref{Eq:xhat}), for any point $\bx^e$ in the hemi-ellipsoid $\Omega$. We also introduce the standard polar angles $\theta$ and $\phi$, given by Equations \ref{Eq:theta} and \ref{Eq:phi}, and illustrated in Figure \ref{fig:cellularMotion}a. A radial velocity vector can then be defined as in Equation (\ref{Eq:transformation}).

    \begin{subequations}
    \begin{align}
    \hat{\bx}^\text{e} &= \frac{\bx^\text{e}}{\|\bx^\text{e}\|} \label{Eq:xhat}\\
    \theta &= {\cos}^{-1}(\hat{x_3}^\text{e}) \label{Eq:theta} \\
    \phi &= {\sin}^{-1}\Big( \frac{\hat{x_1}^\text{e}}{\cos{\theta}} \Big) \label{Eq:phi}\\
    \bv^{\text{r}} &=  {v}^\text{c}\begin{Bmatrix}
    \cos{\phi}\sin{\theta} \\
    \sin{\phi}\sin{\theta} \\
    \cos{\theta} \\
    \end{Bmatrix} \label{Eq:transformation}
    \end{align}
    \end{subequations}
    
Here, $v^\text{c} = \Vert\bv^\text{c}\Vert$ is the magnitude of the radial velocity vector $\bv^\text{c}$ from the spherical geometry, with value given in Table \ref{Tbl:table1}, and $\bv^\text{r}$ is the radial velocity in an ellipsoidal geometry.\footnote{As defined here, $\bv^\text{r} = \bv^\text{c}$. We introduce $\bv^\text{r}$ with the intention of further scaling it; the seemingly superfluous definition of $\bv^\text{r}$ is to keep the construction of the radial field distinct from the spherically symmetric field.} An easy calculation reveals that the sphere-to-ellipsoid transformation biases $\bv^\text{r}$ along the longer axes of the ellipsoid, in comparison with  $\bv^\text{c}$. This is seen to some degree in Figure \ref{fig:cellularMotion}a, where $\bv^\text{r}$ is more aligned with the long axis, an effect that becomes more pronounced with aspect ratio. When used in the partial differential equation (\ref{Eq:cell_motion}), this bias creates a band of high advection velocity, manifesting in increased cell concentration in the plane of the longer axes, shown in Figure \ref{fig:cellularMotion}b. 

Another consequence of the sphere-to-ellipsoid transformation is nonuniform ellipsoidal shell thickness. To account for this effect, we identify points on the inner and outer surfaces, $\bx_\text{min} \in \partial\Omega^\text{i}$ and $\bx_\text{max}  \in \partial\Omega^\text{o}$, along a vector, $\hat{\bx}^\text{e}$ (or $\bv^\text{r}$), as shown in Figure \ref{fig:cellularMotion}a (right) and described in Equations (\ref{Eq:x-max}) and (\ref{Eq:x-min}). There, $r_\text{i}$ and $R_\text{o}$ are the inner and outer spherical radii, which are transformed into the semi-axes of inner and outer ellipsoids. 

We define a position-dependent thickness scaling factor, $t^\text{r}_\text{scale}$, as the difference between $\Vert\bx_{\text{max}}\Vert$ and $\Vert\bx_{\text{min}}\Vert$ divided by the thickness of the sphere, $R_\text{o} - r_\text{i}$ (Equation (\ref{Eq:thicknessa})). It falls to its minimum value in the $\be_2$ direction, as seen in Figure \ref{fig:cellularMotion}a. This scaling is then applied to the mapped radial velocity, $\bv^\text{r}$, to give the final, scaled velocity vector, $\bv^{\text{r}}_\text{scale}$, shown in Equation (\ref{Eq:radial_scaled}). Cell migration described by Equation (\ref{Eq:cell_motion}) is then solved with $\bv= \bv^{\text{r}}_\text{scale}$.

\begin{subequations}
\begin{align}
\bx_{\text{max}} &= \begin{Bmatrix}
\alpha \cdot R_\text{o} \cos{\phi}\sin{\theta} \\
\beta \cdot R_\text{o} \sin{\phi}\sin{\theta} \\
\gamma \cdot R_\text{o} \cos{\theta} \\
\end{Bmatrix} \label{Eq:x-max} \\
\bx_{\text{min}} &= \begin{Bmatrix}
\alpha\cdot r_\text{i} \cos{\phi}\sin{\theta} \\
\beta \cdot r_\text{i} \sin{\phi}\sin{\theta} \\
\gamma \cdot r_\text{i} \cos{\theta} \\
\end{Bmatrix} \label{Eq:x-min} \\
t^\text{r}_{\text{scale}} & = \frac{\|\bx^\text{e}_{\text{max}}\| -\|\bx^\text{e}_{\text{min}}\|}{R_\text{o}-r_\text{i}} \label{Eq:thicknessa}\\
\bv^{\text{r}}_\text{scale} &= t^\text{r}_\text{scale}\bv^{\text{r}}  \label{Eq:radial_scaled}
\end{align}
\end{subequations}
\subsection{Normal advection velocity in the ellipsoidal geometry}
\label{sec:normalmap}
As discussed above, a radial vector field in an ellipsoid is distinct from one that is normal to concentric ellipsoidal surfaces. In order to generate the latter type of field, we begin by computing the ellipsoidal surface, say $\psi$, on which a point, $\bx^\text{e} \in \Omega$, lies. 

\begin{figure}[h]
\includegraphics[scale = 0.15]{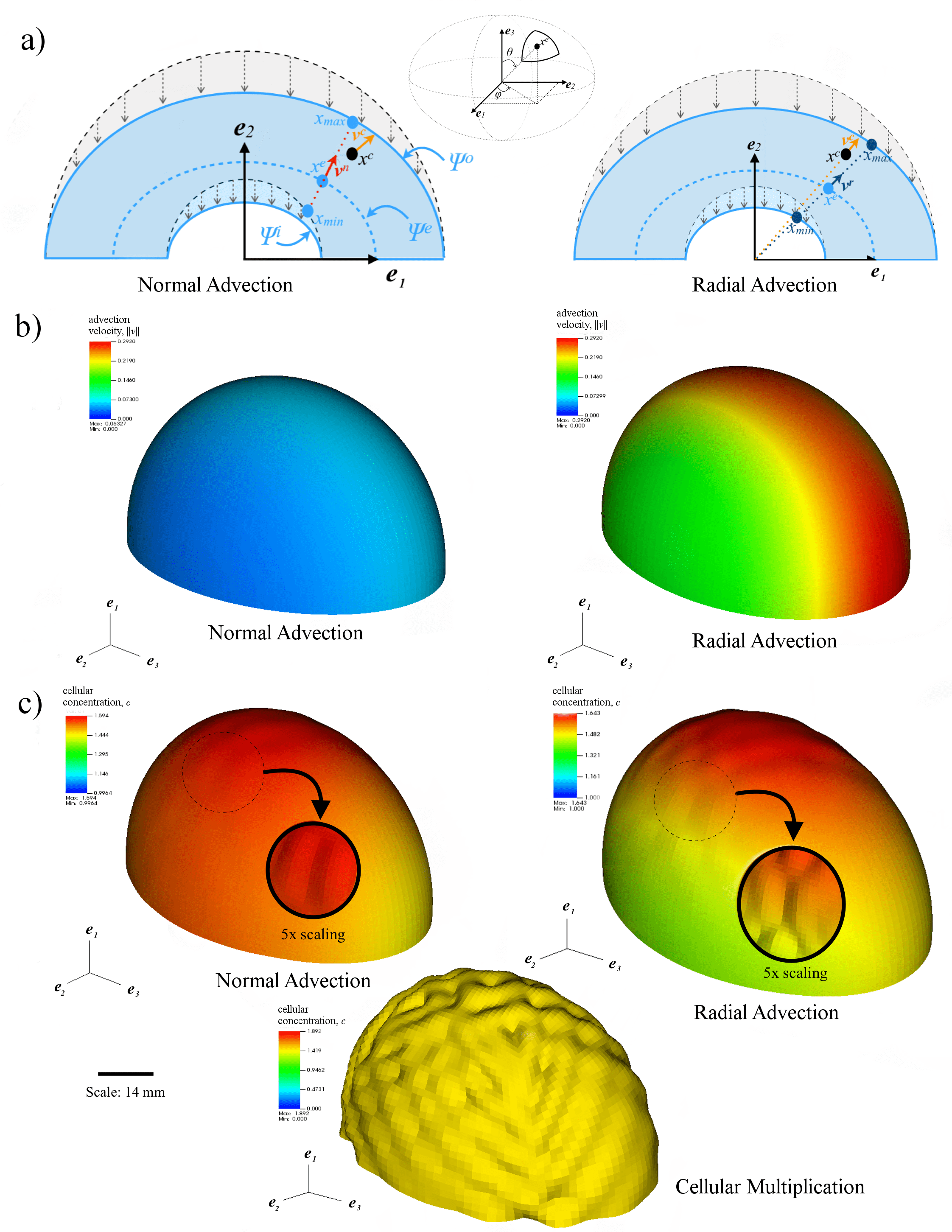}
\centering
\caption{a) Illustration of two mappings of advection velocity b) nonuniform cellular velocities for normal (left) and radial (right) mappings, c) resulting patterning of two mappings of advection (top) and cellular proliferation (bottom). All three models use a 7\% cortical thickness.}
\label{fig:cellularMotion}
\end{figure}

Using the scaling factors $\alpha,\beta,\gamma$ we write the equation of an ellipsoid, obtained by transforming a spherical surface of radius $R^\text{c}$, as
\begin{equation}
\left(\frac{{x_1^\text{e}}}{\alpha R^\text{c}}\right)^2+\left(\frac{{x_2^\text{e}}}{\beta R^\text{c}}\right)^2+\left(\frac{{x_3^\text{e}}}{\gamma R^\text{c}}\right)^2 = 1 \\
\end{equation}
 We introduce Equation (\ref{Eq:etoc}), which maps any point from the hemi-ellipsoid, $\bx^\text{e}$, back to its originating point $\bx^\text{c}$ in the hemisphere. This leads to Equation (\ref{Eq:Rc}) for $R^\text{c}$ in terms of $\bx^\text{e}$.
\begin{subequations}
\begin{align}
\bx^\text{c} &= \begin{Bmatrix} x_1^\text{e}/\alpha\\
x^\text{e}_2/\beta\\
x^\text{e}_3/\gamma \end{Bmatrix} \label{Eq:etoc}\\
R^\text{c} &= \sqrt{\Big( \frac{x_1^\text{e}}{\alpha} \Big)^2+\Big( \frac{x^\text{e}_2}{\beta} \Big)^2+\Big( \frac{x^\text{e}_3}{\gamma} \Big)^2}\label{Eq:Rc} 
\end{align}
\end{subequations}
Then, $\psi(\bx^\text{e}) - 1 = 0$ is the ellipsoidal surface, containing the point $\bx^\text{e}$, where $\psi$ is given by Equation (\ref{Eq:psi}).  We then generate an advection vector field $\bv^\text{n}$ that is everywhere normal to $\psi - 1 = 0$. This is illustrated in Figure \ref{fig:cellularMotion}a (left) and shown in Equation (\ref{Eq:transformation_normal}).
\begin{subequations}
\begin{align}
\psi(\bx^{\text{e}}) &= \left(\frac{x^{\text{e}_1}}{\alpha R^{\text{c}} }\right)^2+\left(\frac{x^{\text{e}}_2}{\beta R^{\text{c}}}\right)^2+\left(\frac{x^{\text{e}_3}}{\gamma R^{\text{c}}}\right)^2 \label{Eq:psi}\\
\bv^{\text{n}} &= {v}^{\text{c}} \frac{\nabla\psi}{\| \nabla \psi \|} \label{Eq:transformation_normal}
\end{align}
\end{subequations}

Similarly to Equation \ref{Eq:radial_scaled}, we scale $\bv^\text{n}$ to generate $\bv^\text{n}_\text{scale}$. We define $\tilde{\bx}^\text{e}_\text{max}$ as the intersection of the line containing $\bx^\text{e}$ and parallel to $\bv^\text{n}$, and the outer hemi-ellipsoidal surface ($R^\text{c} = R^\text{o}$ in Equation (\ref{Eq:psi})). Analogously, define $\tilde{\bx}^\text{e}_\text{min}$ as the intersection of the line containing $\bx^\text{e}$ and parallel to $\bv^\text{n}$, and the inner hemi-ellipsoidal surface ($R^\text{c} = r^\text{i}$ in Equation (\ref{Eq:psi})). The scaling follows as shown in Equations (\ref{Eq:thicknessb}-\ref{Eq:normal_scaled}). Cell migration described by Equation (\ref{Eq:cell_motion}) is then solved with $\bv = \bv^{\text{n}}_\text{scale}$.

\begin{subequations}
\begin{align}
t^\text{n}_{\text{scale}} &= \frac{\|\tilde{\bx}^\text{e}_{\text{max}}\|-\|\tilde{\bx}_{\text{min}}\|}{R_\text{o}-r_\text{i}} \label{Eq:thicknessb}\\
\bv^\text{n}_\text{scale} &= t^\text{r}_\text{scale}\bv^\text{n} \label{Eq:normal_scaled}
\end{align}
\end{subequations}

\subsection{Comparison of the influences of cell velocity fields on folding patterns in the ellipsoidal geometry}
\label{sec:compare}
The description of  neuronal migration in the neurodevelopmental literature \cite{Jin2003,Sun2014,Ronan2014} is not presented in terms that would delineate the cell velocity fields as radial or normal. Therefore, we compared these two migration mechanisms for patterning in the ellipsoidal geometry against the development of the early Central Sulcus in fetal MRI \cite{Gholipour2014,Gholipour2017}. As reference we also included the migration-less, cell proliferation model.

The results appear in Figure \ref{fig:cellularMotion}c. We note that both the radial and normal models of advection velocity give rise to a structure, which on closer examination, resembles an incipient Central Sulcus. (See Conclusions, Section \ref{sec:concl} for a discussion on the extent of invagination of the incipient Central Sulcus in these computations.) This impression is based on its location in the coronal plane, and vertical orientation, both of which features were identified in the Introduction as relevant for defining the Central Sulcus. Both radial and normal models of advection velocity demonstrate an incipient Central Sulcus. While the radial velocity produces a slightly more prominent sulcal structure, more complete computations are needed to follow this structure past the 24 to 25 gestational weeks stages before the suitability of one model over the other can be conclusively established (see the discussion in Conclusions, Section \ref{sec:concl}). On the other hand, the cell proliferation model, which neglects cell advection and diffusion develops a cross latticed pattern without forming an incipient structure resembling the Central Sulcus. We therefore conclude that cell migration, in addition to being observed during neurological development \cite{Sun2014}, is important to the physics of cortical folding and for attaining anatomically relevant morphologies. The radial advection model is carried through to Sections \ref{sec:corticalthick} and \ref{sec:energy} because of its slightly more prominent sulcal structure in Figure \ref{fig:cellularmotion}c.

\section{Interaction of cell migration and geometry: The influence of cortical layer thickness}
\label{sec:corticalthick}
Several cortical folding studies have examined the buckling of a thin, elastic layer on an elastic substrate \cite{Bayly2013,Budday2017,Budday2015_3,Hong2009,Jin2011,Yin2008}. These analytic and computational investigations have considered a variety of geometries including plates and shells to elucidate the role of the layer-to-substrate thickness ratio in forming folding patterns. We therefore consider the implications of cortical thickness for the folding patterns and for developing anatomically accurate structures. Because the cells migrate into the cortical layer and expand tangentially there, this part of the study combines the influence of the cell migration mechanism and geometry.

Using fetal MRI data at 24 gestational weeks (the smooth state before formation of the Central Sulcus) \cite{Habas2012,Gholipour2017}, we found an average cortical thickness of $2$ mm. From the dimensions of the average 24 week-old fetal brain in the same data set, this corresponds to a $7$\% cortical thickness in our model. We conducted a sensitivity study of the folding pattern to cortical thickness, using values of $7, 10, 14$ and $18$ percent to determine this parameter's role in the development of the Central Sulcus. The results appear in Figure \ref{fig:corticalThickness}a, and show three features to the interaction of cell migration with cortical thickness to influence patterning:

\begin{figure}[h]
\includegraphics[scale = 0.15]{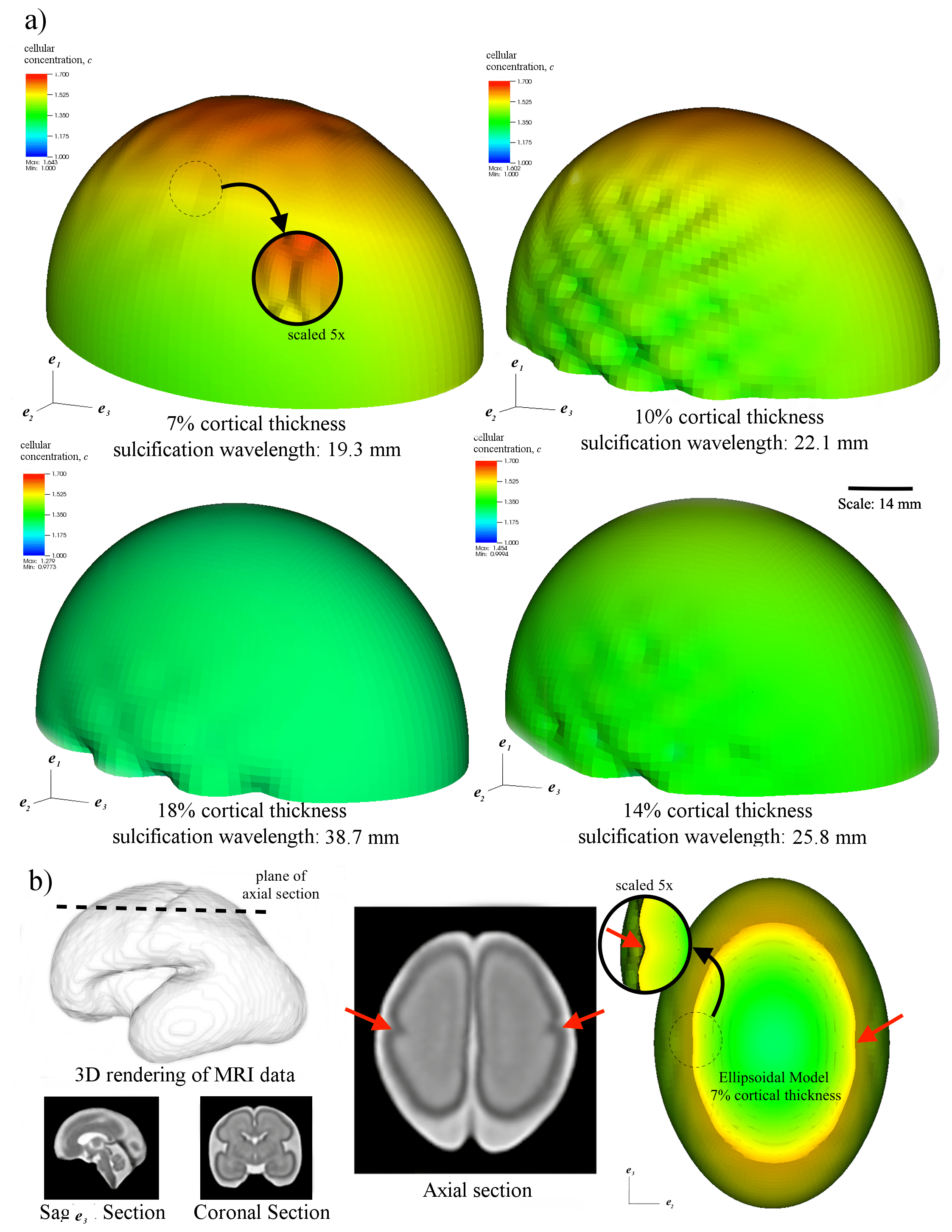}
\centering 
\caption{a) Morphology and sulcification wavelengths resulting from variations of cortical thickness and b) axial cut of $7$\% cortical thickness model (right) as compared to fetal MRI data of a similar section at 25 gestational weeks (left) \cite{Gholipour2014, Gholipour2017}.}
\label{fig:corticalThickness}
\end{figure}
\begin{itemize}
   \item[a)] Sulcification wavelength increases with the cortical shell thickness as follows: 19.3 mm (7\% thickness), 22.1 mm (10\% thickness), 25.8 mm (14\% thickness) and 38.7mm (18\% thickness). This observation corresponds with the key finding in the work of Yin et al \cite{Yin2008}, although that study did not model cell migration. Measurements of wavelengths were taken in the $\be_2-\be_3$ plane, at the locations of deepest sulcification, or $x_1$ values of 28.8 mm (7\% thickness), 21.6 mm (10\% thickness), 7.2 mm (14\% thickness) and 0 mm (18\% thickness). The origin of the coordinate system lies at the center of the basal plane of the computational model shown in Figure \ref{fig:corticalThickness}a.
    \item[b)] The location of sulcification is also sensitive to cortical thickness. Moving in the negative $\be_1$ direction from the crown (in the caudal direction), the transition from smooth to folded cortex occurs further from the pole as the cortical thickness increases. The $x_1$-coordinate at which sulcification begins shifts from $x_1 = 20$ mm (the north pole of the ellipsoid) with 7\% thickness to 18.1 mm, 15.6 mm and 10 mm in the 10, 14 and 18\% thicknesses, respectively.
    \item[c)] The $7$\% cortical thickness case displays a morphology most similar to that seen in fetal MRI data between 24 and 25 gestational weeks \cite{Gholipour2014, Gholipour2017}, namely the development of an incipient Central Sulcus-like structure, circled in Figure \ref{fig:corticalThickness}a.
    \end{itemize}
Figure \ref{fig:corticalThickness}b shows a collection of MR images of the 25 gestational week brain. The two images on the right offer a comparison of the axial section through the Central Sulcus with a corresponding section through the deformed state of the computational model with 7\% cortical thickness. The 25 gestational week Central Sulcus on the MRI, and its incipient counterpart in the computation are pointed out by red arrows.

\noindent\textbf{Remark 1}. The ellipsoidal geometries in Figures \ref{fig:geometry}c (right), \ref{fig:cellularMotion}c, \ref{fig:corticalThickness}a (7\%) and \ref{fig:energy}c demonstrate folding along the intersection of the sagittal plane and dorsal surface. This is not an anatomical feature. A more accurate fetal geometric model would include the longitudinal fissure in this position, and these folds would not develop.
\section{The energy variations that drive folding} 
\label{sec:energy}
The existence of multiple solution paths at a bifurcation implies non-uniqueness, and an associated instability of the system. Post-bifurcation states of deformation displaying folds lie on low energy branches. This furnishes a reason to investigate the total elastic free energy along the solution. In the absence of body force and traction, the elastic free energy is the integral of the neo-Hookean strain energy density function (\ref{Eq:neohookean}) over the domain, as shown in Equation (\ref{Eq:energy}).\\
\begin{equation}
E(t) = \int\limits_{\Omega_0} W (\bF^e(\bX,t)) d\textrm{V}
\label{Eq:energy}
\end{equation}
\indent The boundary conditions in Equation (\ref{Eq:TranspBCa}) and (\ref{Eq:TranspBCb}) create a flux of cells into the cortex, where their accumulation drives growth and elastic deformation. The progressive elastic deformation in cortical regions that are not undergoing folding contributes an increasing energy density that compensates for the decrease in other regions that do fold into lower energy density states. Consequently, $E(t)$ is an increasing function, and does not display decreases associated with bifurcations. However, $t$ represents the loading of the one-parameter function, $E(t)$. The second time derivative of the $\text{d}^2 E/\text{d}t^2$, therefore, is a measure of system stiffness. Of special interest are intervals in which $\text{d}^2 E/\text{d}t^2 < 0$, corresponding to elastic instability, and potential bifurcations. Consider Figure \ref{fig:energy} showing the evolution of $\text{d}^2 E/\text{d}t^2$ with a hemi-ellipsoidal geometry, radial advection velocity, and $7$\% cortical thickness. This is the case leading to the formation of a structure resembling an incipient Central Sulcus (Figures \ref{fig:corticalThickness}a and \ref{fig:corticalThickness}b). The second derivative is computed by a second-order central difference scheme. The curve of $\text{d}^2 E/\text{d}t^2$ is correlated with the evolution of the folded pattern, which is shown at three critical time instants (also see Supplementary Movie 1):

\begin{figure}[h]
\includegraphics[scale = 0.15]{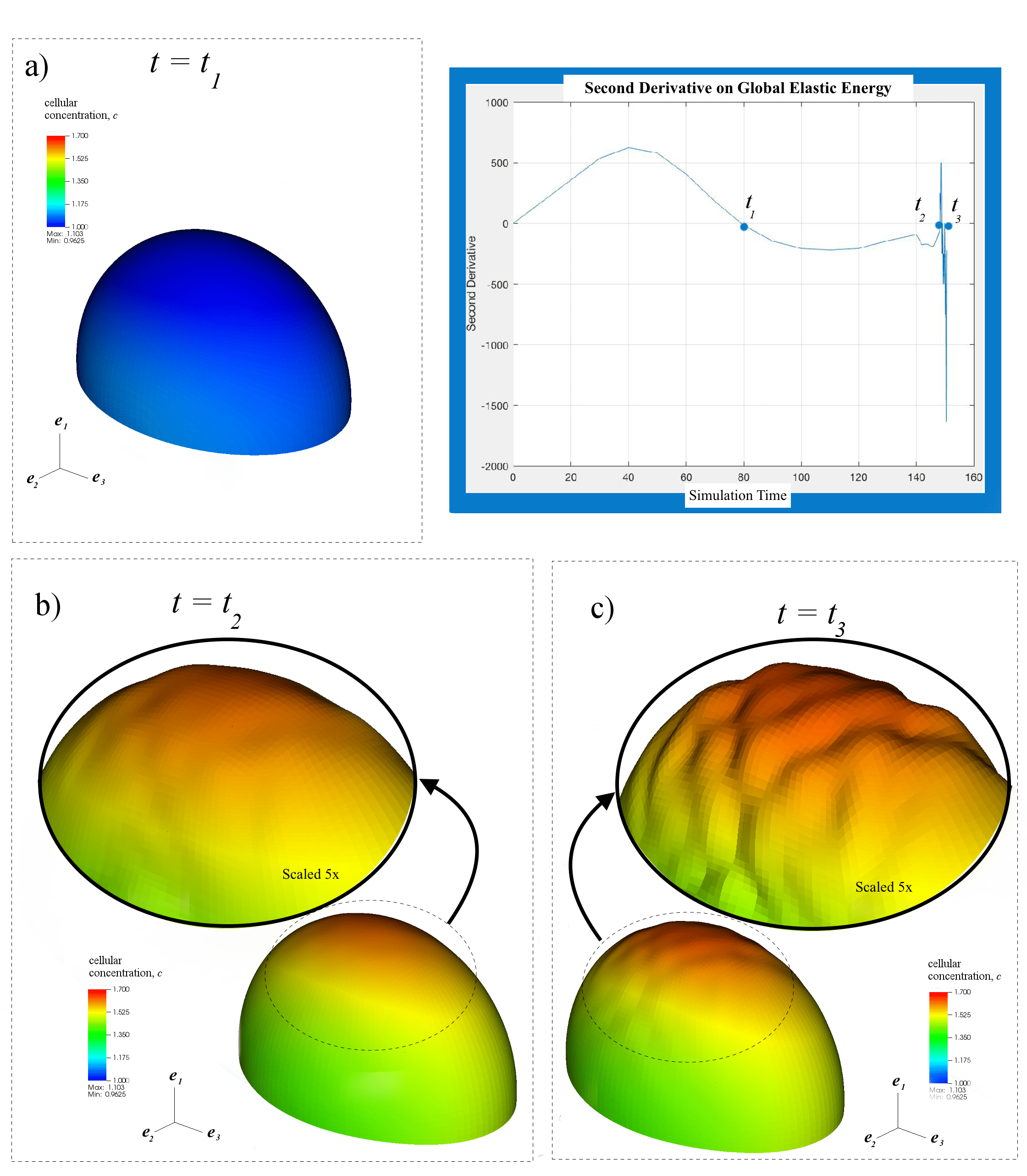}
\centering 
\caption{The stiffness, $\text{d}^2 E/\text{d}t^2$ \emph{versus} $t$, at three critical time instants}.
\label{fig:energy}
\end{figure}

 \begin{itemize}
\item[a)] Figure \ref{fig:energy}a shows the growing brain at time $t_1 = 80$ when $\text{d}^2 E/\text{d}t^2 = 0$, and is decreasing. Because $\text{d}^2 E/\text{d}t^2 \ge 0$ for $t \in [0, t_1]$, the system is elastically stable over this interval, and the deformation does not display a bifurcation from the smooth state.
\item[b)] Figure \ref{fig:energy}b shows the model brain at time $t_2$, chosen such that $\text{d}^2 E/\text{d}t^2 < 0$ for $t \in (t_1, t_2)$. Over this interval of elastic instability, the deformation has undergone a number of bifurcations from the smooth state of deformation, and displays several emerging sulci and gyri.
\item[c)] At times $t \ge t_2$, the stiffness undergoes large fluctuations. The sub-interval with negative stiffness, $\text{d}^2 E/\text{d}t^2 < 0$, is accompanied by the onset of pronounced, as well as more widespread sulcification and gyrification. (The numerical resolution of the second time derivative has degraded, due to which the curve shows an excursion of $\text{d}^2 E/\text{d}t^2$ into the positive half-plane.) This folded state displays an incipient fold in the position corresponding to the Central Sulcus.    
\end{itemize}

This approach offers insight to the onset of bifurcations and the development of post-bifurcation deformation by correlating them with the evolution of the stiffness. The regimes of non-positive stiffness $\text{d}^2 E/\text{d}t^2 \le 0$ during this computation were negotiated by adaptive time-stepping, which controls the extent of cell migration and growth, and therefore the onset of local bifurcations from the smooth state of deformation.
\section{Conclusions}
\label{sec:concl}
This letter initiates a study of normal morphological development of the human brain, specifically of the sequence of sulcification and gyrification.  Our focus is on (a) the physical mechanisms of migration and tangential intercalation of neurons in the cortex, which lead to growth, and (b) the influence of geometry. The mechanisms are governed by the advection-diffusion-reaction equation and inhomogeneous growth in the setting of nonlinear elasticity, respectively. Using these mathematical models, shapes informed by medical imaging data and a finite element framework, we also have identified and studied the influence of overall system geometry and thickness of the cortical layer. 

This letter represents only the first stage in our studies. For this reason, we have focused on the Central Sulcus, one of the first primary folds to develop in the fetal human brain. We are able to make three important conclusions: As discussed in Section \ref{sec:geom}, a hemi-ellipsoidal geometry leads to folded morphologies that include a structure resembling the early Central Sulcus in fetal development. The associated transformation in our study from a hemispherical to the hemi-ellipsoidal geometry suggests at least two models of cell migration, both of which collapse to radial migration on the hemispherical geometry. The influences of these mechanisms of migration are compared for reference with a strict cell proliferation model. In Sections \ref{sec:geom} and \ref{sec:cellmotion}, we explored these treatments and concluded that inclusion of cell migration is important for properly modelling sulcification and gyrification. However, nothing resembling the Central Sulcus forms in the model neglecting migration. This suggests that the gradient of cell concentration induced by the mechanism of migration influences the subsequent distribution of growth strains to create the Central Sulcus-like structure as seen in Figure \ref{fig:cellularMotion}c with the radial and normal advection velocity fields.

In Section \ref{sec:corticalthick}, our model suggests the importance of an anatomically grounded treatment of cortical thickness. Medical imaging data suggests a cortical layer corresponding to a $7$\% thickness in our model. A numerical study of variable cortical thickness confirmed that the value motivated by MRI data indeed produced results bearing similarities to those seen in early Central Sulcus formation.

Finally, in Section \ref{sec:energy}, we discussed the second derivative of the elastic strain energy as a measure of the system's stiffness, elastic stability and as an indicator of bifurcations. We anticipate that this approach will prove important in future studies of the sequence of primary sulcification and gyrification.

The computation with an ellipsoidal geometry, radial advection velocity and 7\% cortical thickness shows what we have termed an \emph{incipient Central Sulcus} in the approximate location and with roughly the orientation of the Central Sulcus that begins to develop between 24 and 25 gestational weeks. Beyond the configuration shown in Figures \ref{fig:corticalThickness}a (upper left) and \ref{fig:corticalThickness}b the elasticity problem fails to converge in our computations. We have not sought to smooth the mesh on this configuration and further drive the computation to a post-bifurcated state with a deeper invagination of this structure. This is mainly because we recognize that our reference ellipsoidal geometry also remains a far from ideal representation of the anatomy of the 24 gestational week brain (Figure \ref{fig:Potato}a, left). Our study, as we have emphasized throughout is a preliminary exploration of the roles of geometric features (overall aspect ratios and cortical thickness), cell migration proliferation mechanisms, and their interaction in establishing the initial folding pattern resembling the targeted Central Sulcus. In future work we will incorporate higher-fidelity data for a systematic study of brain morphology.

\section{Acknowledgements}
SV was supported by an NSF Graduate Fellowship, a Michigan Institute for Computational Discovery and Engineering (MICDE) Fellowship and a University of Michigan Mechanical Engineering Department Fellowship. Computing resources were provided by the University of Michigan's Flux cluster. These forms of support are gratefully acknowledged.

%
%
\bibliographystyle{abbrvnat}
\bibliography{brainFolding}

\end{document}